\documentclass{article}
\usepackage{amssymb}

\input{tcilatex}
\begin{document}

\begin{center}
{\LARGE The Dirac equation in Schwarzschild mass coupled to a Stationary
Electromagnetic Field}

A. Al-Badawi and M.Q. Owaidat

Department of Physics, Al-Hussein Bin Talal University, P. O. Box: 20,
71111, Ma'an, Jordan

Email: ahmadalbadawi@hotmail.com and Owaidat@ahu.edu.jo

{\Large Abstract}
\end{center}

We study the Dirac equation in a spacetime that represents the nonlinear
superposition of the Schwarzchild solution to an external, stationary
electromagnetic Berttoti-Robinson solution. We separate the Dirac equation
into radial and angular equations using Newman--Penrose formalism. We obtain
exact analytical solutions of the angular equations. We manage to obtain the
radial wave equations with effective potentials. Finally, we study the
potentials by plotting them as a function of radial distance and examine the
effect of the twisting parameter and the frequencies on the potentials.

\section{Introduction}

The spacetime we are considering represents the metric of the Schwarzchild
(S) mass coupled to an externally twisting electromagnetic (em) universe,
namely, generalized Berttoti-Robinson (GBR) spacetime [1]. We shall refer to
this solution as the SGBR. This metric is an example to coupling black hole
solutions with the surroundings. Limiting case of this spacetime includes
the case of a stationary em universe i.e. Berttoti-Robinson (BR) geometry.
Recall that the BR solution [2-3], describes a non-singular, spherically
symmetric universe filled with a static electric field. It is considered as
a unique conformally flat solution of the source-free Einstein-Maxwell
equations for non-null fields. This spacetime is the direct product of
2-sphere and 2-dimensional anti-de Sitter spacetime. Therefore BR solution
has the symmetries of both these spaces with 6 parameter isometry group. In
this present paper our interesting metric represents a black hole which is
embedded into an em field. Similar properties of our interesting metric was
found long time ago by Carter [4]. In contrast, the metric in [4] is given
in terms of BR coordinates which are valid inside the throat region only.
But, this metric uses S coordinates that refers to the entire universe.

Studying Dirac equation in different backgrounds has been extensively worked
out and published in literature. For example, in S geometry [5-6] and Kerr
spacetime [7-10]. BR geometry [11-12], where in [11] Silva-Ortigoza showed
how the Dirac equation could be separated in the BR spacetime with
cosmological constant. Whereas in [12], a more detailed study on the problem
of the Dirac equation in the BR background was worked out. Dirac equation in
rotating BR\ geometry [13], they have obtained exact solutions to both the
radial equations and the angular massless neutrino equations. Where the
exact solution of the radial equations is given in terms of hypergeometric
functions. Recently, Dirac equation was examined in Kerr-Taub-NUT spacetime
[14]. In the last reference, using Boyer-Lindquist coordinates they manage
to separate the Dirac equation into radial and angular parts. They obtained
\ exact solution of the angular equations for some special cases and
obtained the radial wave equations with an effective potentials and exposed
the effect of the NUT parameter. Other similar studies with different
background, Nutku helicoid spacetime [15], Kerr--Newman--AdS black hole
geometry [16] and in the background of the Kerr--Newman family, which has
been considered in several studies [17-18].

Here, we study the solution of the Dirac equation in the S mass coupled to a
GBR spacetime. The set of equations representing the Dirac equation in the
Newman--Penrose (NP) formalism is decoupled into a radial (function of
distance $r$ only) and an angular parts (function of angle $\theta $ only).
The angular equation is solved exactly in terms of associated Legendre
functions. The radial equations are discussed and the one dimensional Schr%
\"{o}dinger-like wave equations with effective potentials are obtained.
Finally in order to understand and expose the effect of the twisting
parameter and frequencies on the potentials, curves are plotted and
discussed.

Our paper is organized as follows: in section 2, we present the Dirac
equation in the metric of SGBR\ solution and separate them into two parts.
In section 3, solutions of the angular and radial equations are presented.
Finally, we discuss and comment on the effect of the twist parameter and the
frequency on the potentials by plotting them as a function of radial
distance.

\section{Dirac equation in the metric of SGBR}

The metric we are dealing with represents the non-linear superposition of
the S solution and the GBR solution[1], is given by

\begin{equation}
ds^{2}=\frac{r^{2}-2Mr}{r^{2}f(r)}\left[ dt-Mq(1+a^{2})\cos \theta d\phi %
\right] ^{2}-\frac{r^{2}f(r)}{r^{2}-2Mr}dr^{2}-r^{2}f(r)\left( d\theta
^{2}+\sin ^{2}\theta d\phi ^{2}\right) .
\end{equation}

where 
\begin{equation}
r^{2}f(r)=\frac{1}{2}r\left( r-2M\right) \left[ p\left( 1+a^{2}\right)
+a^{2}-1\right] +2Mar+M^{2}\left[ p\left( 1+a^{2}\right) -2a\right]
\end{equation}%
in which, $M$ is the S mass, $p=$ $\sqrt{1+q^{2}}$ is the twisting parameter
of the external em field and $0\leq a\leq 1$ is the interpolation parameter.
It is seen that for $a=0\left( q=0,p=1\right) $ metric (1) reduces to BR
solution and for $a=1\left( q=0,p=1\right) $ it reduces to S solution. One
can justify the limiting cases by calculating the NP curvatures using the
null tetrad \ 1-form of NP formalism. we make the choice of the following
null tetrad basis 1-forms $\left( l,n,m,\overline{m}\right) $ of the NP
formalism [19] in terms of the metric functions that satisfies the
orthogonality conditions, $\left( l.n=-m\text{.}\overline{m}=1\right) $.
Where a bar over a quantity denotes complex conjugation. We can write the
covariant 1-forms as

\[
\sqrt{2}l=H(r)(dt-Q\cos \theta d\phi )-\frac{dr}{H(r)}, 
\]%
\[
\sqrt{2}n=H(r)(dt-Q\cos \theta d\phi )+\frac{dr}{H(r)}, 
\]%
\ 
\begin{equation}
\sqrt{2}m=r\sqrt{f\left( r\right) }(d\theta +i\sin \theta d\phi ),
\end{equation}

where 
\begin{equation}
H^{2}(r)=\frac{r^{2}-2Mr}{r^{2}f\left( r\right) },\qquad Q=M\sqrt{p^{2}-1}%
\left( 1+a^{2}\right) .
\end{equation}%
Using the above null tetrad (3), we obtain the non-zero Weyl and Ricci NP
scalars [1]%
\begin{eqnarray*}
\Psi _{2} &=&\frac{-M}{4R^{6}}[2aa_{0}r^{3}+Ma_{0}\left( 4c_{0}-b_{0}\right)
r^{2}+2M^{2}\left( b_{0}^{2}-b_{0}c_{0}-3a_{0}c_{0}\right) r \\
&&+2M^{3}b_{0}c_{0}+iq\left( 1+a^{2}\right) \left( a_{0}\left( 3M-r\right)
r^{2}+2M^{2}\left( 1-a^{2}\right) r-2c_{0}M^{3}\right) ],
\end{eqnarray*}%
\begin{equation}
\Phi _{11}=\frac{M^{2}\left( 1-a^{2}\right) }{2R^{4}},
\end{equation}%
where 
\begin{eqnarray}
R^{2} &=&r^{2}f\left( r\right) , \\
c_{0} &=&p\left( 1+a^{2}\right) -2a,  \nonumber \\
b_{0} &=&a_{0}-2a,  \nonumber \\
a_{0} &=&p\left( 1+a^{2}\right) +a^{2}-1.  \nonumber
\end{eqnarray}

This metric is a type-D metric whose conformal curvature is due to the twist
parameter $p.$ It is seen that metric (1) satisfies all the required limits
as boundary conditions:

\begin{eqnarray}
\ BR\ Solution &\rightarrow &\left( a=0,q=0,M=1\right) \rightarrow \Psi
_{2}=0,\Phi _{11}=\frac{1}{2}, \\
\ S\,Solution &\rightarrow &\left( a=1,q=0\right) \rightarrow \Psi _{2}=%
\frac{-M}{r^{3}},\Phi _{11}=0,  \nonumber \\
GBR\ Solution &\rightarrow &\left( a=0,q\neq 0\right) \rightarrow \Psi
_{2}\neq 0\neq \Phi _{11}.  \nonumber
\end{eqnarray}

The case $GBR\ Solution\rightarrow \left( a=0,q\neq 0\right) $ has been
studied in greater detail in [20], where its metric is an exact solution of
Einstein-Maxwell equations of general relativity which represents a uniform
em field in the rotating state.

We shall now write the Dirac equation in the SGBR spacetime. The Dirac
equation in the NP formalism [8] is given by%
\[
\left( D+\epsilon -\rho \right) F_{1}+\left( \overline{\delta }+\pi -\alpha
\right) F_{2}-i\mu _{0}G_{1}=0, 
\]%
\[
\left( \Delta +\mu -\gamma \right) F_{2}+\left( \delta +\beta -\tau \right)
F_{1}-i\mu _{0}G_{2}=0, 
\]%
\[
\left( D+\overline{\epsilon }-\overline{\rho }\right) G_{2}-\left( \delta +%
\overline{\pi }-\overline{\alpha }\right) G_{1}-i\mu _{0}F_{2}=0, 
\]%
\begin{equation}
\left( \Delta +\overline{\mu }-\overline{\gamma }\right) G_{1}-\left( 
\overline{\delta }+\overline{\beta }-\overline{\overline{\tau }}\right)
G_{2}-i\mu _{0}F_{1}=0,
\end{equation}

where $\mu _{0}$ is the mass of the Dirac particle.

Let us write the corresponding directional derivatives as

\[
\sqrt{2}D=\frac{1}{H}\partial t+H\partial r, 
\]%
\[
\sqrt{2}\Delta =\frac{1}{H}\partial t-H\partial r, 
\]%
\[
\sqrt{2}\delta =-\frac{1}{R}\left[ \partial _{\theta }+i\frac{1}{\sin \theta 
}\partial _{\phi }\right] , 
\]%
\begin{equation}
\sqrt{2}\bar{\delta}=-\frac{1}{R}\left[ \partial _{\theta }-i\frac{1}{\sin
\theta }\partial _{\phi }\right] ,
\end{equation}

Using the above tetrad we determine the nonzero NP complex spin coefficients
[19] as, 
\[
\rho =\mu =\frac{-R^{\prime }H}{\sqrt{2}R}+i\frac{HQ}{2\sqrt{2}R^{2}}, 
\]

\[
\epsilon =\gamma =\frac{H^{\backprime }}{2\sqrt{2}}+i\frac{HQ}{4\sqrt{2}R^{2}%
}, 
\]

\begin{equation}
\alpha =-\beta =\frac{\cot \theta }{2\sqrt{2}R}.
\end{equation}

The form of the Dirac equation suggests that we assume [8], 
\begin{equation}
F_{1}=F_{1}\left( r,\theta \right) e^{i\left( kt+n\phi \right) },  \nonumber
\end{equation}%
\[
F_{2}=F_{2}\left( r,\theta \right) e^{i\left( kt+n\phi \right) }, 
\]%
\[
G_{1}=G_{1}\left( r,\theta \right) e^{i\left( kt+n\phi \right) }, 
\]%
\begin{equation}
G_{2}=G_{2}\left( r,\theta \right) e^{i\left( kt+n\phi \right) }.
\end{equation}%
We consider the corresponding Compton wave of the Dirac particle as in the
form of $F=F\left( r,\theta \right) e^{i\left( kt+n\phi \right) }$ , where $%
k $ is the frequency of the incoming wave and $n$ is the azimuthal quantum
number of the wave. Substituting the appropriate spin coefficients (10) and
the spinors (11) into the Dirac equation (8) , we obtain

\[
\left( H\mathbf{D}-i\frac{HQ}{4R^{2}}\right) F_{1}-\frac{1}{R}\mathbf{L}%
F_{2}-i\mu _{0}G_{1}=0, 
\]%
\[
\left( -H\mathbf{D}^{\dag }+i\frac{HQ}{4R^{2}}\right) F_{2}-\frac{1}{R}%
\mathbf{L}^{\dag }F_{1}-i\mu _{0}G_{2}=0, 
\]%
\[
\left( H\mathbf{D}+i\frac{HQ}{4R^{2}}\right) G_{2}+\frac{1}{R}\mathbf{L}%
^{\dag }G_{1}-i\mu _{0}F_{2}=0, 
\]%
\begin{equation}
\left( -H\mathbf{D}^{\dag }-i\frac{HQ}{4R^{2}}\right) G_{1}+\frac{1}{R}%
\mathbf{L}G_{2}-i\mu _{0}F_{1}=0,
\end{equation}%
where the radial and the angular operators are 
\begin{eqnarray}
\mathbf{D} &\mathbf{=}&\frac{d}{dr}+\frac{H^{\prime }}{2H}+\frac{R^{\prime }%
}{R}+i\frac{k}{H^{2}} \\
\mathbf{D}^{\dag } &=&\frac{d}{dr}+\frac{H^{\prime }}{2H}+\frac{R^{\prime }}{%
R}-i\frac{k}{H^{2}}  \nonumber \\
\mathbf{L} &\mathbf{=}&\frac{d}{d\theta }+\frac{n}{\sin \theta }+\frac{\cot
\theta }{2}  \nonumber \\
\mathbf{L}^{\dag } &=&\frac{d}{d\theta }-\frac{n}{\sin \theta }+\frac{\cot
\theta }{2}  \nonumber
\end{eqnarray}%
It is now apparent that Eqs. (13) can be separated by implying the
separability ansatz%
\begin{eqnarray}
F_{1} &=&f_{1}\left( r\right) A_{1}\left( \theta \right) , \\
F_{2} &=&f_{2}\left( r\right) A_{2}\left( \theta \right) ,  \nonumber \\
G_{1} &=&f_{2}\left( r\right) A_{1}\left( \theta \right) ,  \nonumber \\
G_{2} &=&f_{1}\left( r\right) A_{2}\left( \theta \right) .  \nonumber
\end{eqnarray}%
With this ansatz, Eqs. (13) become%
\[
\left[ \left( RH\mathbf{D}-i\frac{HQ}{4R}\right) f_{1}-i\mu _{0}Rf_{2}\right]
A_{1}-\left[ \mathbf{L}A_{2}\right] f_{2}=0, 
\]%
\[
\left[ \left( -RH\mathbf{D}^{\dag }+i\frac{HQ}{4R}\right) f_{2}-i\mu
_{0}Rf_{1}\right] A_{2}-\left[ \mathbf{L}^{\dag }A_{1}\right] f_{1}=0, 
\]%
\[
\left[ \left( RH\mathbf{D}+i\frac{HQ}{4R}\right) f_{1}-i\mu _{0}Rf_{2}\right]
A_{2}+\left[ \mathbf{L}^{\dag }A_{1}\right] f_{2}=0, 
\]%
\begin{equation}
\left[ \left( -RH\mathbf{D}^{\dag }-i\frac{HQ}{4R}\right) f_{2}-i\mu
_{0}Rf_{1}\right] A_{1}+\left[ \mathbf{L}A_{2}\right] f_{1}=0.
\end{equation}%
These equations (15) imply that%
\[
\mathbf{L}A_{2}=\lambda _{1}A_{1}, 
\]%
\[
\mathbf{L}^{\dag }A_{1}=\lambda _{2}A_{2}, 
\]%
\[
\mathbf{L}^{\dag }A_{1}=\lambda _{3}A_{2}, 
\]%
\begin{equation}
\mathbf{L}A_{2}=\lambda _{4}A_{1},
\end{equation}
\[
\left( RH\mathbf{D}-i\frac{HQ}{4R}\right) f_{1}-i\mu _{0}Rf_{2}=\lambda
_{1}f_{2}, 
\]%
\[
\left( -RH\mathbf{D}^{\dag }+i\frac{HQ}{4R}\right) f_{2}-i\mu
_{0}Rf_{1}=\lambda _{2}f_{1}, 
\]%
\[
\left( RH\mathbf{D}+i\frac{HQ}{4R}\right) f_{1}-i\mu _{0}Rf_{2}=-\lambda
_{3}f_{2}, 
\]%
\begin{equation}
\left( -RH\mathbf{D}^{\dag }-i\frac{HQ}{4R}\right) f_{2}-i\mu
_{0}Rf_{1}=-\lambda _{4}f_{1},
\end{equation}%
where $\lambda _{1},$ $\lambda _{2},$ $\lambda _{3}$ and $\lambda _{4}$ are
four constants of separation. However, let us assume $\left( \lambda
_{1}=\lambda _{4}=\lambda ,\lambda _{2}=\lambda _{3}=-\lambda \right) ,$ and
adding equations, then we obtain the radial and the angular pair equations

\begin{equation}
RH\left( \mathbf{D}f_{1}\right) =\left( \lambda +i\mu _{0}R\right) f_{2},
\end{equation}%
\begin{equation}
RH\left( \mathbf{D}^{\dag }f_{2}\right) =\left( \lambda -i\mu _{0}R\right)
f_{1},
\end{equation}%
\begin{equation}
\mathbf{L}A_{2}=\lambda A_{1},
\end{equation}%
\begin{equation}
\mathbf{L}^{\dag }A_{1}=-\lambda A_{2}.
\end{equation}

\section{Solution of angular and radial equations}

$\qquad $Angular Eqs. (20) and (21) we can be written as

\begin{equation}
\frac{dA_{1}}{d\theta }+\left( \frac{\cot \theta }{2}-\frac{n}{\sin \theta }%
\right) A_{1}=-\lambda A_{2},
\end{equation}%
\begin{equation}
\frac{dA_{2}}{d\theta }+\left( \frac{\cot \theta }{2}+\frac{n}{\sin \theta }%
\right) A_{2}=\lambda A_{1}.
\end{equation}%
The structure of the angular equations admits the solution of $A_{1}$ is
similar to $A_{2}$. Thus, it is enough to decouple the angular equations for 
$A_{1}$.

First, we affect the transformation

\[
A_{1}\left( \theta \right) =\cos \left( \frac{\theta }{2}\right) S_{1}+\sin
\left( \frac{\theta }{2}\right) S_{2} 
\]

\begin{equation}
A_{2}\left( \theta \right) =-\sin \left( \frac{\theta }{2}\right) S_{1}+\cos
\left( \frac{\theta }{2}\right) S_{2}
\end{equation}

Then, let $x=\cos \theta $ and one can write Eq. (22) into second order
differential equation

\begin{equation}
\left( 1-x^{2}\right) \frac{d^{2}S_{1}}{dx^{2}}-2x\frac{dS_{1}}{dx}+\left[
\lambda \left( \lambda +1\right) -\frac{\left( n+\frac{1}{2}\right) ^{2}}{%
1-x^{2}}\right] S_{1}=0,
\end{equation}

with $\left( n+\frac{1}{2}\right) ^{2}\leq \lambda ^{2}$. The solution can
be obtained in terms of the associated Legendre functions

\begin{equation}
P_{\lambda }^{n}\left( x\right) =\left( 1-x^{2}\right) ^{\tau /2}\frac{d^{n}%
}{dx^{n}}P_{\lambda }\left( x\right) .
\end{equation}

For the radial equations (18) and (19) can be rearranged as

\begin{equation}
\frac{df_{1}}{dr}+\left( \frac{R^{2}HH^{\prime }}{2H^{2}R^{2}}+\frac{%
RR^{\prime }H^{2}}{H^{2}R^{2}}+i\frac{kR^{2}}{H^{2}R^{2}}\right) f_{1}=\frac{%
1}{RH}\left( i\mu _{0}R+\lambda \right) f_{2},
\end{equation}

\begin{equation}
\frac{df_{2}}{dr}+\left( \frac{R^{2}HH^{\prime }}{2H^{2}R^{2}}+\frac{%
RR^{\prime }H^{2}}{H^{2}R^{2}}-i\frac{kR^{2}}{H^{2}R^{2}}\right) f_{2}=\frac{%
1}{RH}\left( \lambda -i\mu _{0}R\right) f_{1}.
\end{equation}%
Our aim now is to put the radial equations (27) and (28) in the form of one
dimensional wave equations and obtain the effective potentials. Therefore,
to achieve our aim we follow the method applied by Chandrasekhar's book [8].
We will starts with the transformations

\begin{equation}
P_{1}=RHf_{1},\qquad P_{2}=RHf_{2}.
\end{equation}%
and next the scaling 
\begin{equation}
T_{1}=P_{1}\exp \left( \dint \frac{H^{\prime }}{2H}dr\right) ,\qquad
T_{2}=P_{2}\exp \left( \dint \frac{H^{\prime }}{2H}dr\right)
\end{equation}%
Then, Eqs. (27) and (28) become%
\begin{equation}
\frac{dT_{1}}{dr}+i\frac{kR^{2}}{H^{2}R^{2}}T_{1}=\frac{1}{RH}\left( i\mu
_{0}R+\lambda \right) T_{2},
\end{equation}%
\begin{equation}
\frac{dT_{2}}{dr}-i\frac{kR^{2}}{H^{2}R^{2}}T_{2}=\frac{1}{RH}\left( \lambda
-i\mu _{0}R\right) T_{1}.
\end{equation}%
Assume 
\begin{equation}
\frac{du}{dr}=\frac{1}{H^{2}}.
\end{equation}%
Therefore, the above equations (31) and (32) in terms of the new independent
variable $u,$ become%
\begin{equation}
\frac{dT_{1}}{du}+ikT_{1}=\frac{H}{R}\left( i\mu _{0}R+\lambda \right)
T_{2},\qquad
\end{equation}%
\begin{equation}
\frac{dT_{2}}{du}-ikT_{2}=\frac{H}{R}\left( \lambda -i\mu _{0}R\right) T_{1}.
\end{equation}%
where 
\begin{eqnarray}
u &=&\frac{1}{2}\left[ p\left( 1+a^{2}\right) +a^{2}-1\right] r+\frac{1}{2}%
\sqrt{\frac{M}{2r}}(\left[ p\left( 1+a^{2}\right) +a^{2}-4a-1\right] r 
\nonumber \\
&&-\left[ p\left( 1+a^{2}\right) -2a\right] )M\arctan \left( \sqrt{\frac{r}{%
2M}}\right) .
\end{eqnarray}%
Let us apply another transformation, namely the new functions \ 
\begin{equation}
T_{1}=\phi _{1}\exp \left( \frac{-i}{2}\arctan \left( \frac{\mu _{0}R}{%
\lambda }\right) \right) ,\qquad T_{2}=\phi _{2}\exp \left( \frac{i}{2}%
\arctan \left( \frac{\mu _{0}R}{\lambda }\right) \right) ,
\end{equation}%
with these transformations, Eqs. (34) and (35) take the form%
\begin{equation}
\frac{d\phi _{1}}{du}+ik\left( 1-\frac{H^{2}}{2k}\left[ \frac{\mu
_{0}\lambda R^{\prime }}{\lambda ^{2}+\mu _{0}^{2}R^{2}}\right] \right) \phi
_{1}=\frac{H}{R}\sqrt{\lambda ^{2}+\mu _{0}^{2}R^{2}}\phi _{2},\qquad
\end{equation}%
\begin{equation}
\frac{d\phi _{2}}{du}-ik\left( 1-\frac{H^{2}}{2k}\left[ \frac{\mu
_{0}\lambda R^{\prime }}{\lambda ^{2}+\mu _{0}^{2}R^{2}}\right] \right) \phi
_{2}=\frac{H}{R}\sqrt{\lambda ^{2}+\mu _{0}^{2}R^{2}}\phi _{1}.
\end{equation}%
Again change the variable $u$ into $\widehat{r}$ as 
\begin{equation}
\widehat{r}=u-\frac{1}{2k}\arctan \left( \frac{\mu _{0}R}{\lambda }\right) .
\end{equation}%
Then we can write Eqs.(38) and (39) in the alternative forms 
\begin{equation}
\frac{d\phi _{1}}{d\widehat{r}}+ik\phi _{1}=W\phi _{2},
\end{equation}%
\begin{equation}
\frac{d\phi _{2}}{d\widehat{r}}-ik\phi _{2}=W\phi _{1},
\end{equation}%
where 
\begin{equation}
W=\frac{2k\left( \lambda ^{2}+\mu _{0}^{2}R^{2}\right) ^{3/2}H}{2k\left(
\lambda ^{2}+\mu _{0}^{2}R^{2}\right) R-\lambda \mu _{0}R^{\prime }RH^{2}}.
\end{equation}%
Finally, in order to put the above equations (41) and (42) into one
dimensional Schr\"{o}dinger-like wave equations, we define%
\begin{equation}
2\phi _{1}=\psi _{1}+\psi _{2},\qquad 2\phi _{2}=\psi _{1}-\psi _{2}.
\end{equation}%
Then Eqs.(41) and (42) become%
\[
\frac{d\psi _{1}}{d\widehat{r}}-W\psi _{1}=-ik\psi _{2}, 
\]%
\begin{equation}
\frac{d\psi _{2}}{d\widehat{r}}+W\psi _{2}=-ik\psi _{1}.
\end{equation}%
Which can be cast into 
\[
\frac{d^{2}\psi _{1}}{d\widehat{r}^{2}}+k^{2}\psi _{1}=V_{+}\psi _{1}, 
\]%
\begin{equation}
\frac{d^{2}\psi _{2}}{d\widehat{r}^{2}}+k^{2}\psi _{2}=V_{-}\psi _{2},
\end{equation}%
where the effective potentials can be obtained from 
\begin{equation}
V_{\pm }=W^{2}\pm \frac{dW}{d\widehat{r}}.
\end{equation}%
We calculate the potentials as%
\begin{eqnarray}
V_{\pm } &=&\frac{4k^{2}L^{3/2}RH}{D^{2}}[L^{3/2}RH\pm 3\mu
_{0}^{2}R^{3}R^{\prime }H^{2}\pm L\left( R^{\prime }H+RH^{\prime }\right) 
\nonumber \\
&&\mp \frac{2kLR^{2}H^{2}}{I}\left( 2\mu _{0}^{2}R^{3}R^{\prime }-\frac{%
\lambda \mu _{0}}{2k}B+\frac{2RR^{\prime }}{k}L\right) ],
\end{eqnarray}%
where%
\begin{eqnarray}
L &=&\left( \lambda ^{2}+\mu _{0}^{2}R^{2}\right) ,\qquad
D=2kL^{2}R^{2}-\lambda \mu _{0}R^{\prime }R^{2}H^{2},\qquad  \nonumber \\
B &=&R^{\prime }R^{2}H^{2}+2RHR^{\prime }\left( R^{\prime }H+RH^{\prime
}\right) .
\end{eqnarray}%
Let us note that for the case of massless Dirac particle $\mu _{0}=0$, the
potentials take the form%
\begin{equation}
V_{\pm }=\frac{\lambda H}{R^{3}}\left[ \lambda RH\pm \left( R^{\prime
}H+RH^{\prime }\right) \mp 2\frac{H^{2}RR^{\prime }}{k}\right] .
\end{equation}%
Now we will find the complete solution of Eq. (46). First we can rewrite Eq.
(46) as%
\[
\frac{d^{2}\psi _{1}}{d\widehat{r}^{2}}+\left( k^{2}-V_{+}\right) \psi
_{1}=0, 
\]%
\begin{equation}
\frac{d^{2}\psi _{2}}{d\widehat{r}^{2}}+\left( k^{2}-V_{-}\right) \psi
_{2}=0.
\end{equation}%
This is nothing but the one dimensional Schr\"{o}dinger wave equations with
total energy of the wave $k^{2}$. Eq. (51) can be solved by WKB\
approximation method [21,22]. The solution is given by 
\[
\psi _{1}=\sqrt{T_{1}\left[ \omega _{1}\left( \widehat{r}\right) \right] }%
e^{iy_{1}}+\sqrt{R_{1}\left[ \omega _{1}\left( \widehat{r}\right) \right] }%
e^{-iy_{1}}, 
\]%
\begin{equation}
\psi _{2}=\sqrt{T_{2}\left[ \omega _{2}\left( \widehat{r}\right) \right] }%
e^{iy_{2}}+\sqrt{R_{2}\left[ \omega _{2}\left( \widehat{r}\right) \right] }%
e^{-iy_{2}},
\end{equation}%
where 
\begin{equation}
\omega _{1}\left( \widehat{r}\right) =\sqrt{\left( k^{2}-V_{+}\right) }%
,\qquad \omega _{2}\left( \widehat{r}\right) =\sqrt{\left(
k^{2}-V_{-}\right) }
\end{equation}%
\begin{equation}
y_{1}\left( \widehat{r}\right) =\int \omega _{1}\left( \widehat{r}\right) d%
\widehat{r}+\text{constant},\qquad y_{2}\left( \widehat{r}\right) =\int
\omega _{2}\left( \widehat{r}\right) d\widehat{r}+\text{constant}
\end{equation}%
with 
\begin{equation}
T_{1}\left( r\right) +R_{1}\left( r\right) =1,\qquad T_{2}\left( r\right)
+R_{2}\left( r\right) =1\qquad \text{instantaneously.}
\end{equation}%
Where $\omega $ is the wave number of the incoming wave and $y$ is the 
\textit{eikonal. }$T_{1,2}$ and $R_{1,2}$ are the instantaneous transmission
and reflection coefficients [5] respectively. Let us note that the above
solution is valid when $\left( 1/\omega \right) \left( d\omega /d\widehat{r}%
\right) \ll \omega $. \bigskip

\section{Discussion}

In this section, we are going to expose the effect of the twisting parameter 
$p$ on the effective potentials by plotting the potentials as a function of
radial distance with varying $p$. Next, we will study the behavior of the
potentials by drawing potentials curves for some different value of
frequency $k$. From the potentials Eqs. (48) and (50), it is very clear that
the potentials depend strictly on the twisting parameter $p$ and the
interpolation parameter $a$ via the functions $R$ and $H$. We must notice
that the expressions becomes singular when $D=0$ for the case $\mu _{0}\neq
0,$ and when $R=0$ for the case $\mu _{0}=0$. Also we notice that the
potentials have local extrema when $\frac{dV_{\pm }}{dr}=0,$ which is a very
complicated equation to be solved since the potentials are rather involved.
\ 

Recall that the twisting parameter $p$ has been discussed in details [23],
indeed they interpret it as the twist parameter of the em universe. If there
is no em field this parameter represents the twist of the vacuum spacetime.
Therefore, in this case it couples to S mass and generates a NUT parameter
given by $l=\pm M\sqrt{p-\frac{1}{p}},\left( p>1\right) .$ To examine the
effect of $p$ on the potentials we obtain two dimensional plots of Eq. (48)
for the case $\mu _{0}\neq 0$. Let us note that, the physical regions for
our potentials start from $r>2M$ in all figures. In fig. 1, we exhibit the
behaviour of the potentials $V_{\pm }$ for different values of twisting
parameter $p,$ the case $\left( p=1,\,q=0\right) $ is excluded because it
lead us to a special case in the metric, where we have chosen the rest mass
to be $\mu _{0}=0.12,$ and fixed value of $k=0.2$ such that $\mu _{0}<k$. It
is seen from fig.1 that, potentials have sharp peaks in the physical
distance of $r,$ and when the twisting parameter $p$ increases the values of
the sharp peaks decrease. We conclude that for large value of $p,$ a massive
Dirac particle moving in the physical region faces a small potential
barriers whereas for small value of $p$ encounters large potentials
barriers. Therefore, as the twisting parameter increases the kinetic energy
of the particle increases and advances without strong retarding potentials.
Anyhow, potentials are bounded, regardless of the value of $p$, which means
that as $\left( r\rightarrow \infty \right) ,$ the potentials $V_{\pm
}\rightarrow $ constant as seen from all figures. To examine the behaviour
of the potentials for some values of frequencies we obtain figure 2 for
various value of $k$ and fixed value of $p=10$. It is seen from fig. 2 that,
the general behavior of \ the potentials are not changed, they still have
sharp peaks and behave similar for large distances. However, we notice that
at low frequencies the sharp peaks are very clear, but for high frequencies
the levels of sharp peaks decrease. In the case of massless Dirac particle,
we would remark that in the two dimensional plots the behavior of the
potentials are similar to the case $\mu _{0}\neq 0.$

Now, the three dimensional plot of the potentials with respect to the
twisting parameter and the radial distance $r$ is given in Figure 3. It
shows the effect of the twisting parameter on the the potentials explicitly
for the massive Dirac particle $\left( \mu _{0}=0.12\text{ and }k=0.2\right)
.$ It is seen from Fig. 3 that, high potential barriers are observed for
small value of twisting parameter whereas for small value of $p$ potential
barriers decrease. Again for large distances, potentials levels decrease and
asymptote behaviour is manifested. Note that, the behaviour of the $V_{+},$
which is given in figure 4, is similar as in Figure 3. The three dimensional
plots of potentials with respect to frequency, for fixed twisting parameter $%
\left( p=10\text{ and }\mu _{0}=0.12\right) $, and the radial distance is
given in figure 5. We observe \ in Fig. 5 that, sharp peaks are clear for
low frequencies only. Now, for massless Dirac particle $\left( \mu
_{0}=0\right) ,$ three dimensional plots are given in figures 6 and 7. The
general behaviour of the potentials for massless Dirac particle shows
similar behavior as for massive case. However, the change of the sharp peaks
and the levels for potential barriers as the twisting parameter increases is
weak compared with the massive case. Therefore, the influence of the
twisting parameter on the massless case is not strong as in the massive one.

\section{Conclusion}

In this paper, we have considered the Dirac equation in a background that
represents the coupling of S mass and the GBR spacetime, using NP\ null
tetrad formalism. By employing an axially symmetric ansatz for the Dirac
spinor, we managed to separate the equation into radial and angular parts.
We were able to obtain an exact solution for the angular equations in terms
of associated Legendre functions. We studied the radial equations and
obtained the radial wave equations with effective potentials. Finally, we
studied the behaviour of the potentials by varying the twisting parameter
and the frequency, by plotting two and three dimensional plots. We showed
that, as the twisting parameter increases, the potentials barriers decrease.
However, for low frequencies sharp peaks of potentials are more observed
than high frequencies. In the case of massless Dirac particle, it is seen
that the effect of twisting parameter is weaker than the massive one. Our
main motivation was to give an analytical expression of the solution which
could be useful for further study of the thermodynamical properties of the
spinor field in SGBR background. For future work, the study of the charged
Dirac particles in this spacetime may reveal more information compared to
the present case.\medskip

{\Large References\medskip }

[1] Halilsoy M and Al-Badawi A1995 Class. Quantum Grav. \textbf{12} 3013.

[2] Robinson I 1959 Bull. Acad. Pol. Sci. \textbf{7},351.

[3] Bertotti B 1959 Phys. Rev. \textbf{116} 1331.

[4] Carter B 1973 Black Hole Equilibrium States. In: De Witt, C. M., De
Witt. B S(eds) Black Hole pp 57-217 Gordon and Breach New York.

[5] Mukhopadhyay B and Chakrabarti S K 1999 Class. Quantum Grav. \textbf{16}
3165.

[6] M. Halil 1981Int. J. Theor. Phys. \textbf{20} 911.

[7] Chandrasekhar S 1976 Proc. R. Soc. Lond. A \textbf{349 }571.

[8] Chandrasekhar S 1983 The Mathematical Theory of Black Holes Clarendon,
London.

[9] Dolan S R and Gair J R 2009 Class. Quantum Grav. \textbf{26} 175020.

[10] Sakalli I and M Halilsoy 2004 Phys. Rev. D \textbf{69}, 124012.

[11] Silva-Ortigoza G 2001 Gen. Relativ. Gravit. \textbf{5}, 395.

[12] Sakalli I 2003 Gen. Relativ. Gravit. \textbf{35}, 1321.

[13] Al-Badawi A and Sakalli I 2008 J. Math. Phys. \textbf{49 }052501.

[14] Cebeci H and \H{O}zdemir N 2013 Class. Quantum Grav. \textbf{30} 175005.

[15] Birkandan T and Horta\c{c}su M 2007 J. Math. Phys. \textbf{48} 092301.

[16] Belgiorno F and Cacciatori S L 2010 J. Math. Phys. \textbf{51} 033517.

[17] Finster F, Smoller J and Yau S T 2000 J. Math. Phys. \textbf{41} 2173.

[18] Winklmeier M and Yamadan O 2006 J. Math. Phys. \textbf{47} 102503.

[19] Newman E Tand Penrose R 1962 J. Math. Phys. \textbf{3} 566.

[20] Al-Badawi A and Halilsoy M 2004 IL Nuovo Cimento \textbf{119 B }N.10.

[21] Davydov A M 1976 Quantum Mechanics 2nd edn (Oxford: Pergamon).

[22] Mathews J andWalker R L 1970 Mathematical Methods Of Physics 2nd edn
(New York: Benjamin-Cummings).

[23] Al-Badawi A and Halilsoy M 2006 Gen. Relativ. Gravit. \textbf{38}, 1729.

\bigskip

\bigskip

\bigskip

\end{document}